\documentclass[12pt]{article}
\usepackage{a4wide}
\usepackage{epsfig}
\usepackage{amsmath}

\newlength{\absize}
\catcode`@=11
\def\citer{\@ifnextchar [{\@tempswatrue\@citexr}{\@tempswafalse\@citexr[]}}

\def\@citexr[#1]#2{\if@filesw\immediate
  \write\@auxout{\string\citation{#2}}\fi
  \def\@citea{}\@cite{\@for\@citeb:=#2\do
    {\@citea\def\@citea{--\penalty\@m}\@ifundefined
       {b@\@citeb}{{\bf ?}\@warning
       {Citation `\@citeb' on page \thepage \space undefined}}%
\hbox{\csname b@\@citeb\endcsname}}}{#1}}
\catcode`@=12

\begin{document}
  \thispagestyle{empty}
  \pagestyle{empty}
  \renewcommand{\thefootnote}{\fnsymbol{footnote}}
\newpage\normalsize
    \pagestyle{plain}
    \setlength{\baselineskip}{4ex}\par
    \setcounter{footnote}{0}
    \renewcommand{\thefootnote}{\arabic{footnote}}
\newcommand{\preprint}[1]{%
  \begin{flushright}
    \setlength{\baselineskip}{3ex} #1
  \end{flushright}}
\renewcommand{\title}[1]{%
  \begin{center}
    \LARGE #1
  \end{center}\par}
\renewcommand{\author}[1]{%
  \vspace{2ex}
  {\Large
   \begin{center}
     \setlength{\baselineskip}{3ex} #1 \par
   \end{center}}}
\renewcommand{\thanks}[1]{\footnote{#1}}
\begin{flushright}
\end{flushright}
\vskip 0.5cm

\begin{center}
{\large \bf Testing Spatial Noncommutativity via Rydberg Atoms}
\end{center}
\vspace{1cm}
\begin{center}
Jian-zu Zhang$\;^{\ast}$
\end{center}
\vspace{1cm}
\begin{center}
Institute for Theoretical Physics, Box 316, East China University
of Science and Technology, Shanghai 200237, P. R. China
\end{center}
\vspace{1cm}

\begin{abstract}
The possibility of testing spatial noncommutativity via Rydberg
atoms is explored. An atomic dipole of a cold Rydberg atom is
arranged in appropriate electric and magnetic field, so that the
motion of the dipole is constrained to be planar and rotationally
symmetric. Spatial noncommutativity leads to that the canonical
angular momentum possesses fractional values. In the limit of
vanishing kinetic energy, the dominate value of the lowest
canonical angular momentum takes $\hbar/2$. Furthermore, in the
limit of eliminating magnetic field, the dominate value of the
lowest canonical angular momentum changes from $\hbar/2$ to
$\hbar/4$. This result is a clear signal of spatial
noncommutativity. An experimental verification of this prediction
is suggested.

\end{abstract}

\begin{flushleft}
$^{\ast}$ E-mail address: jzzhang@ecust.edu.cn

\end{flushleft}
\clearpage

In hinting new physics in the present round it seems that physics
in noncommutative space \citer{CDS} is a candidate. This is
motivated by studies of low energy effective theory of D-brane
with a nonzero NS-NS $B$ field background. Effects of spatial
noncommutativity are only apparent near the string scale, thus we
need to work at a level of noncommutative quantum field theory.
But it is expected that some low energy relics of such effects may
be verified by nowadays experiments, and some phenomenological low
energy effects may be explored in solvable models at a level of
quantum mechanics in noncommutative space(NCQM). In literature
NCQM have been extensively studied \citer{CST,NP}.

In this paper we study the possibility of testing spatial
noncommutativity via Rydberg atoms at the level of NCQM. In
\cite{Baxt,JZZ} it is demonstrated that cold Rydberg atoms play an
interesting role of realizing analogs of Chern-Simons theory
\cite{DJT,DJT1}. An atomic dipole of a cold Rydberg atom is
arranged in appropriate electric and magnetic field, so that the
motion of the dipole is constrained to be planar and rotationally
symmetric. In this case the R\"ontgen term of the Hamiltonian
takes on a form of a Chern-Simons interaction. This term makes
interesting contribution to dynamics. Furthermore, in an
appropriate optical trapping field the elimination of dipole's
kinetic energy could be achieved physically, and the canonical
angular momentum spectrum changes from integers to positive half
integers. An experimental verification of the Chern-Simons feature
of the angular momentum is allowed.

In noncommutative space new features appear. Spatial
noncommutativity leads to that the canonical angular momentum
spectrum possesses fractional values \citer{JZZ03}. In the limit
of vanishing kinetic energy the dominate value of the lowest
canonical angular momentum takes $\hbar/2$. Spatial
noncommutativity permits a further limiting process of eliminating
magnetic field that the dominate value of the lowest canonical
angular momentum changes from $\hbar/2$ to $\hbar/4$. This result
is a clear signal of spatial noncommutativity. A possibility of
testing spatial noncommutativity via cold Rydberg atoms is
suggested.

{\bf NCQM Algebra.} In order to develop the NCQM formulation we
need to specify the phase space and the Hilbert space on which
operators act. The Hilbert space can consistently be taken to be
exactly the same as the Hilbert space of the corresponding
commutative system \citer{CST}.

As for the phase space we consider both space-space
noncommutativity (space-time noncommutativity is not considered)
and momentum-momentum noncommutativity. There are different types
of noncommutative theories, for example, see a review paper
\cite{DN}.

In the case of simultaneously space-space
 noncommutativity and momentum-momentum noncommutativity the
consistent NCQM algebra \cite{JZZ03} are:
\begin{equation}
\label{Eq:xp} [\hat x_{i},\hat x_{j}]=i\xi^2\theta\epsilon_{ij},
\qquad [\hat x_{i},\hat p_{j}]=i\hbar\delta_{ij}, \qquad [\hat
p_{i},\hat p_{j}]=i\xi^2\eta\epsilon_{ij},\;(i,j=1,2)
\end{equation}
where $\theta$ and $\eta$ are the constant parameters, independent
of position and momentum; $\epsilon_{ij}$ is an antisymmetric unit
tensor, $\epsilon_{12}=-\epsilon_{21}=1,$
$\epsilon_{11}=\epsilon_{22}=0;$ The scaling factor
$\xi=(1+\theta\eta/4\hbar^2)^{-1/2}.$

The Hamiltonian of a Rydberg atom in electric and magnetic field
\cite{Baxt,JZZ} is (summation convention is used):
\begin{eqnarray}
\label{Eq:CS-H} \hat H=\frac{1}{2\mu}(\hat
p_i+\frac{1}{2}g\epsilon_{ij}\hat x_j)^2 +\frac{1}{2}\kappa\hat
x_i^2 =\frac{1}{2\mu}\hat p_i^2 +\frac{1}{2\mu}g\epsilon_{ij}\hat
p_i\hat x_j+\frac{1}{2}\mu\omega^2\hat x_i^2,
\end{eqnarray}
where the co-ordinates $\hat x_i$ refer to the laboratory frame of
the Rydberg atom.
\footnote{\; The Rydberg atom is treated as a structureless dipole
moment. In reality it has the internal atomic structure. For the
following discussions effects of the internal structure are
extremely small, and hence can be forgotten.}
The parameter $\mu$ is the mass of the atom. The electric field
$\vec{E}$ acts radially in the $x-y$ plane, $E_i= -\mathcal{E}\hat
x_i,$ $(i=1, 2),$ where $\mathcal{E}$ is a constant, (i=1, 2), and
the constant magnetic field $\vec{B}$ aligns the $z$ axis. The
constant parameters $g=2qB$ and $\kappa=2q\mathcal{E},$ $q(>0)$ is
dipole's electric charge.
\footnote{\; Rydberg atoms are sensitive to external electric
fields. Even Relatively modest electric fields may ionize Rydberg
atoms. If we ignore the Stark shift of a Rydberg state of
effective principal quantum number $n^{\ast}$ we obtain the
classical electric field $E_c$ for ionization
$E_c=(16{n^{\ast}}^4)^{-1}\; a. u.=3.2\cdot(n^{\ast})^{-4}\cdot
10^8\;V/cm$. In our discussion the electric field $E_i$ should
satisfy the condition $|E|<E_c$.}
The term $g\epsilon_{ij}\hat p_i\hat x_j/(2\mu)$ takes the
Chern-Simons interaction. The frequency
$\omega=\left[g^2/(4\mu^2)+\kappa/\mu \right]^{1/2},$ where the
dispersive ``mass'' term $g/(2\mu)$ comes from the presence of the
Chern-Simons term.

The motivation of considering both space-space and
momentum-momentum noncommutativity is as follows. There are
different ways to construct creation-annihilation operators. We
first construct the deformed annihilation-creation operators
$(\hat a_i$, $\hat a_i^\dagger)$ $(i=1,2)$ at the non-perturbation
level which are related to the variables $(\hat x_i, \hat p_i)$:
\begin{equation}
\label{Eq:aa+1} \hat a_i=\sqrt{\frac{\mu\omega}{2\hbar}}\left
(\hat x_i +\frac{i}{\mu\omega}\hat p_i\right).
\end{equation}
(\ref{Eq:aa+1}) and the NCQM algebra (\ref{Eq:xp}) show that the
operators $\hat a_{i}^\dagger$ and $\hat a_{j}^\dagger$ for the
case $i\ne j$ do not commute. When the state vector space of
identical bosons is constructed by generalizing one-particle
quantum mechanics, because of such a noncommutativity  the
operators $\hat a_1^\dagger\hat a_2^\dagger$ and $\hat
a_2^\dagger\hat a_1^\dagger$ applied successively to the vacuum
state $|0\rangle$ do not produce the same physical state. In order
to maintain Bose-Einstein statistics at the non-perturbation level
described by $\hat a_i^\dagger$ the basic assumption is that
operators $\hat a_i^\dagger$ and $\hat a_j^\dagger$ should be
commuting. This requirement leads to a consistency condition of
NCQM algebra
\begin{equation}
\label{Eq:dd} \eta=\mu^2\omega^2 \theta.
\end{equation}
From (\ref{Eq:xp}), (\ref{Eq:aa+1}) and (\ref{Eq:dd}) it
follows that the commutation relations of $\hat a_i$ and $\hat
a_j^\dagger$ read
\begin{equation}
\label{Eq:[a,a+]1} [\hat a_1,\hat a_1^\dagger]=[\hat a_2,\hat
a_2^\dagger]=1, [\hat a_1,\hat a_2]=0;\quad [\hat a_1,\hat
a_2^\dagger] =i\xi^2\mu\omega \theta/\hbar.
\end{equation}
The first three equations in (\ref{Eq:[a,a+]1}) are the same
commutation relations as the one in commutative space.

The last equation in (\ref{Eq:[a,a+]1}) codes effects of spatial
noncommutativity. We emphasize that it is consistent with all
principles of quantum mechanics and Bose-Einstein statistics.

If momentum-momentum is
commuting, $\eta= 0$,
we could not obtain the third equation in
(\ref{Eq:[a,a+]1}). It is clear that in order to maintain
Bose-Einstein statistics for identical bosons at the level of
$\hat a_i$ and $\hat a_i^\dagger$ we should consider both
space-space noncommutativity and momentum-momentum
noncommutativity.

Now we consider perturbation expansions of $(\hat x_i, \hat p_j)$
and $(\hat a_i, \hat a_j^\dagger).$
The NCQM algebra (\ref{Eq:xp}) has different perturbation
realizations \cite{NP}. We consider the following consistent
ansatz of the perturbation expansions of $\hat x_{i}$ and $\hat
p_{i}$
\begin{equation}
\label{Eq:hat-x-x} \hat
x_{i}=\xi[x_{i}-\theta\epsilon_{ij}p_{j}/(2\hbar)
 ], \quad \hat
p_{i}=\xi[p_{i}+\eta\epsilon_{ij}x_{j}/(2\hbar)].
\end{equation}
where $[x_{i},x_{j}]=[p_{i},p_{j}]=0,
[x_{i},p_{j}]=i\hbar\delta_{ij}.$ In commutative space the
relations between the variables $(x_i, p_i)$ and the
annihilation-creation operators $(a_i, a_i^\dagger)$ are
$a_i=\sqrt{\mu\omega/(2\hbar)}[x_i + ip_i/(\mu\omega)],$
where $[a_{i},a_{j}]=[a_i^\dagger,a_j^\dagger]=0,
[a_{i},a^{\dagger}_{j}]=\delta_{ij}.$ Inserting these relations
into (\ref{Eq:hat-x-x}), using (\ref{Eq:dd}) and comparing the
results with (\ref{Eq:aa+1}), we obtain the perturbation
expansions of $\hat a_i$ and $\hat a_i^\dagger$
\begin{equation}
\label{Eq:hat-a-a1} \hat a_{i}=\xi[a_{i}+i\mu\omega\theta
\epsilon_{ij}a_j/(2\hbar)],\quad
\hat a_{i}^\dagger=\xi[a_{i}^\dagger-i\mu\omega
\theta\epsilon_{ij}a_j^\dagger/(2\hbar)].
\end{equation}

(\ref{Eq:xp}) and (\ref{Eq:aa+1})-(\ref{Eq:hat-a-a1}) are
consistent each other.

{\bf Spectrum 0f Rydberg Atoms.} As in commutative space the
angular momentum is defined as an exterior product $\hat
J=\epsilon_{ij}\hat x_i\hat p_j$. From (\ref{Eq:xp}) and
(\ref{Eq:dd}) it follows that $[\hat J,\hat H]=0.$ Thus $\hat H,
\hat J$ constitute a complete set of observables of the system.

In the following our attention is focused on the perturbation
investigation of $\hat H$ and $\hat J.$ Using
(\ref{Eq:hat-x-x}) we obtain
\begin{eqnarray}
\label{Eq:CS-H1} \hat H=\frac{1}{2M}(
p_i+\frac{1}{2}G\epsilon_{ij} x_j)^2 +\frac{1}{2}K x_i^2
=\frac{1}{2M} p_i^2+\frac{1}{2M}G\epsilon_{ij} p_i
x_j+\frac{1}{2}M\Omega^2 x_i^2,
\end{eqnarray}
where the effective parameters $M, G, K$ and $\Omega$ are defined
as
$1/(2M)\equiv \xi^2\left[c_1^2/(2\mu) +\kappa \bar
\theta^{\;2}/2\right]$,
$G/(2M)\equiv \xi^2\left(c_1 c_2/\mu +\kappa \bar \theta\right)$,
$M\Omega^2/2\equiv \xi^2\left[c_2^2/(2\mu) +\kappa/2 \right]$,
$K\equiv M\Omega^2- G^2/(4M)$,
and $c_1=1+g\bar \theta/2,\;c_2=g/2+\bar \eta,\;\bar
\theta=\theta/(2\hbar),\;\bar \eta=\eta/(2\hbar).$

(\ref{Eq:CS-H1}) is exactly solvable \cite{Baxt,JZZ}. We introduce
new variables $(X_{\alpha}, P_{\alpha})$,
$X_a=\sqrt{M\Omega/(2\omega_a)}x_1-\sqrt{1/(2M\Omega\omega_a)}p_2$,
$X_b=\sqrt{M\Omega/(2\omega_b)}x_1+\sqrt{1/(2M\Omega\omega_b)}p_2$,
$P_a=\sqrt{\omega_a/(2M\Omega)}p_1+\sqrt{M\Omega\omega_a/2}x_2$,
$P_b=\sqrt{\omega_b/(2M\Omega)}p_1-\sqrt{M\Omega\omega_b/2}x_2$,
where $\omega_a=\Omega+G/(2M),\quad \omega_b=\Omega-G/(2M),$ and
define new annihilation operators
$A_{\alpha}=\sqrt{\omega_{\alpha}/(2\hbar)}X_{\alpha}+
i\sqrt{\hbar/(2\omega_{\alpha})}P_{\alpha}$,
$(\alpha=a,b).$ Then the Hamiltonian (\ref{Eq:CS-H1}) decomposes
into two uncoupled harmonic oscillators of unit mass and
frequencies $\omega_a$ and $\omega_b:$
\begin{equation}
\label{Eq:Ha,b} \hat H=H_{a}+H_{b},\quad
H_{\alpha}=\hbar\omega_{\alpha}(A_{\alpha}^{\dagger}A_{\alpha}+
1/2),\quad (\alpha=a,b)
\end{equation}

By a similar procedure we obtain the perturbation expansion of
$\hat J$
\begin{eqnarray}
\label{Eq:Ja,b} \hat J&=&\epsilon_{ij}x_i p_j-\xi^2\left(\bar
\theta p_ip_i +\bar \eta x_ix_i\right)
\nonumber \\
&=&\hbar\left(A_{b}^{\dagger}A_{b}-A_{a}^{\dagger}A_{a}\right)
-\left(A_{a}^{\dagger}A_{a}+A_{b}^{\dagger}A_{b}
+1\right)\mathcal{J}_0, \quad \mathcal{J}_0=\xi^2\mu\omega
\theta\hbar,
\end{eqnarray}
where the zero-point angular momentum $\mathcal{J}_0=\langle
0|J|0\rangle$
codes effects of spatial noncommutativity. It worth noting that it
takes fractional value. The consistency condition (\ref{Eq:dd}) of
NCQM algebra is crucial in the derivation of (\ref{Eq:Ja,b}). The
second line of (\ref{Eq:Ja,b}) is derived by using a relation
$M\Omega=\mu\omega$ which is obtained from (\ref{Eq:dd}).

{\bf Dynamics in the limit of Vanishing Kinetic Energy.} In the
limit of vanishing kinetic energy, $E_k\to 0,$ the Hamiltonian
(\ref{Eq:CS-H1}) shows non-trivial dynamics. In this limit there
are constraints which should be carefully considered. For this
purpose it is more convenient to work in the Lagrangian formulism.
First we identify the limit of vanishing kinetic energy in the
Hamiltonian with the limit of the mass $M\to 0$  in the
Lagrangian. In (\ref{Eq:CS-H1}) in the limit of vanishing kinetic
energy, $\frac{1}{2M}\left( p_i+\frac{1}{2}G\epsilon_{ij}
x_j\right)^2=\frac{1}{2}M \dot{x_i} \dot{x_i}\to 0,$ the
Hamiltonian $H$ reduces to $H_0=\frac{1}{2}K x_i x_i.$ The
Lagrangian corresponding to the Hamiltonian (\ref{Eq:CS-H1}) is
$L=\frac{1}{2}M\dot{x_i}\dot{x_i}
+\frac{1}{2}G\epsilon_{ij}x_i\dot{x_j}-\frac{1}{2}K x_i x_i$. In
the limit of $M\to 0$ this Lagrangian reduces to
$L_0=\frac{1}{2}G\epsilon_{ij}x_i\dot{x_j} -\frac{1}{2}K x_i x_i.$
From $L_0$ the corresponding canonical momentum is
$p_{0i}=\partial L_0/\partial
\dot{x_i}=\frac{1}{2}G\epsilon_{ji}x_j,$ and the corresponding
Hamiltonian is $H_0^{\prime}=p_{0i}\dot{x_i}-L_0=\frac{1}{2}K x_i
x_i=H_0.$ Thus we identify the two limiting processes. We
emphasize when the potential is velocity dependent the limit of
vanishing kinetic energy in the Hamiltonian does not corresponds
to the limit of vanishing velocity in the Lagrangian. If the
velocity approached zero in the Lagrangian there would be no
dynamics. The Hamiltonian (\ref{Eq:CS-H}) and its massless limit
have been studied by Dunne, Jackiw and Trugenberger \cite{DJT1}.

The first equation of (\ref{Eq:CS-H1}) shows that in the limit
$E_k\to 0$ there are constraints
\footnote{\;In this example the symplectic method \cite{FJ} leads
to the same results as the Dirac method for constrained
quantization, and the representation of the symplectic method is
much streamlined.}
\begin{equation}
\label{Eq:Ci} C_i=p_i+\frac{1}{2}G\epsilon_{ij} x_j=0.
\end{equation}
The Poisson brackets of constraints (\ref{Eq:Ci}) are
$\{C_i,C_j\}_P=G\epsilon_{ij}\ne 0,$ so that the corresponding
Dirac brackets of canonical variables $x_i, p_j$ can be determined
\cite{MZ}, $\{x_1,p_1\}_D=\{x_2,p_2\}_D=1/2,\;
\{x_1,x_2\}_D=-1/G,\;\{p_1,p_2\}_D=-G/4$. The Dirac brackets of
$C_i$  with any variables $x_i$  and $p_j$ are zero that the
constraints (\ref{Eq:Ci}) are strong conditions and can be used to
eliminate the dependent variables. If we select $x_1$  and $p_1$
as the independent variables, from (\ref{Eq:Ci}) we obtain
$x_2=-2p_1/G,\;p_2=Gx_1/2.$ The above Dirac brackets show that the
corresponding quantization condition of the independent variables
$x_1$  and $p_1$ is $[x_1,p_1]=i\hbar/2.$ In order to rewrite
$H_0$ in the traditional form we introduce new variables
$q=\sqrt{2}x_1$  and $p=\sqrt{2}p_1,$ which satisfy the normal
quantization condition $[q,p]=i\hbar.$ We introduce the effective
mass $\mu^{\ast}\equiv G^2/2K$ and effective frequency
$\omega^{\ast} \equiv K/G,$ and rewrite the Hamiltonian $H_0$ as
$H_0=\frac{1}{2\mu^{\ast}}p^2+\frac{1}{2}\mu^{\ast}\omega^{\ast 2
}q^2$. Then we define a new annihilation operator
$A=
\sqrt{\mu^{\ast}\omega^{\ast}/2\hbar}\;q
+i\sqrt{\hbar/2\mu^{\ast}\omega^{\ast}}\;p,$
and rewrite the Hamiltonian $H_0$ as
$H_0=\hbar\omega^{\ast}\left(A^\dagger A+1/2\right)$.
Similarly, we rewrite the angular momentum $\hat J$ in
(\ref{Eq:Ja,b}) as
$J_0=\hbar\mathcal{J}_0^{\ast}\left(A^\dagger A+1/2\right)$,
where
$\mathcal{J}_0^{\ast}=1-\xi^2\left(G\bar \theta/2\hbar+2\bar
\eta/G\hbar\right)$.
The eigenvalues of $H_0$ and $J_0$ are, respectively,
$E_n^{\ast}=\hbar\omega^{\ast}\left(n+1/2\right)$,
$\mathcal{J}_n^{\ast}=\hbar\mathcal{J}_0^{\ast}\left(n+1/2\right)$,
$(n=1, 2, \cdots)$. In the limit case $E_k\to 0$ the corresponding
lowest angular momentum is $\hbar\mathcal{J}_0^{\ast}/2$ whose
dominate value is $\hbar/2.$

Because of spatial noncommutativity a further limiting process of
diminishing magnetic field also leads to non-trivial dynamics. In
this limit the parameter $g\to 0,$ the frequency $\omega\to
\omega_0\equiv \sqrt{\kappa/\mu},$ the consistency condition
(\ref{Eq:dd}) is rewritten as
$\bar \eta=\mu^2\omega_0^2 \bar \theta,$ and
$\xi\to\xi_0=(1+\mu^2\omega_0^2\bar \theta^2)^{-1/2}.$
The effective parameters $M, \Omega, G$ and $K$ reduce,
respectively, to the following effective parameters $\tilde M,
\tilde \Omega, \tilde G$ and $\tilde K,$ which are defined by
$\tilde M\equiv [\xi^2_0\left(1/\mu +\kappa \bar
\theta^{\;2}\right)]^{-1}=\mu$,
$\tilde \Omega^2\equiv \xi^2_0\left(\kappa/\mu+\bar
\eta^{\;2}/\mu^2 \right)=\omega_0^2$,
$\tilde G\equiv 2\xi^2_0\left(\mu\kappa \bar \theta+\bar \eta
\right)=2\xi^2_0\mu\kappa \theta/\hbar$
and $\tilde K\equiv \tilde M\tilde \Omega^2- \tilde G^2/(4\tilde
M)=\kappa\left(1-\xi^4_0\mu\kappa \theta^2/\hbar^2\right)$.
Thus in this limit $H_0$ and $J_0$ reduce, respectively, to the
following $\tilde H_0$ and $\tilde J_0:$
\begin{eqnarray}
\label{Eq:H-J02} \tilde H_0&=&\hbar\tilde \omega(\tilde A^\dagger
\tilde A+1/2),
\nonumber\\
\tilde J_0&=&\hbar\mathcal{\tilde J}_0(\tilde A^\dagger \tilde
A+1/2),\quad
\mathcal{\tilde J}_0=1-\xi^2_0(\tilde G\bar \theta/2+2\bar
\eta/\tilde G)=(1-\xi^4_0\mu\kappa \theta^2/\hbar^2)/2,
\end{eqnarray}
where the annihilation operator
$\tilde A=\sqrt{\tilde \mu\tilde \omega/2\hbar}\;q
+i\sqrt{\hbar/2\tilde \mu\tilde \omega}\;p,$
the effective mass $\tilde \mu\equiv \tilde G^2/(2\tilde K)$ and
frequency $\tilde \omega \equiv \tilde K/\tilde G$. From
(\ref{Eq:H-J02}) we conclude that the dominate value of the lowest
angular momentum $\hbar\mathcal{\tilde J}_0/2$ is $\hbar/4.$
Unlike the term $2\xi^2\bar \eta/(G\hbar)\sim \mu\kappa
\theta/(g\hbar^2)\sim 0$ in $\mathcal{J}_0^{\ast}$, here the term
$2\xi^2_0\bar \eta/\tilde G=1/2$ in $\mathcal{\tilde J}_0.$ This
leads to the difference between the dominate values of
$\mathcal{J}_0^{\ast}$ and $\mathcal{\tilde J}_0.$ If we define
the angular momentum with scalar terms $\bar\eta\hat x_i \hat
x_i+\bar\theta\hat p_i \hat p_i$ as in \cite{NP}, we obtain the
same conclusion. This dominate value $\hbar/4$ of the lowest
angular momentum explores the essential new feature of spatial
noncommutativity.
\footnote{\;In the limit of vanishing magnetic field the
Hamiltonian of this system reduces to the Hamiltonian of a
harmonic oscillator. In commutative space the dynamics of a
harmonic oscillator in the limit of vanishing kinetic energy does
not possesses similar constraints. As the kinetic energy
decreased, the potential energy decreases, so that the oscillation
gets weeker and weeker. Thus in commutative space in both limits
of vanishing kinetic energy and magnetic field there is no
dynamics.}

{\bf Testing Spatial Noncommutativity via Rydberg Atoms.}
Following \cite{Baxt,JZZ},
we arrange a cold Rydberg atom in the electric and magnetic fields
with the above suggested arrangement. Assume that the atomic
dipole confined in a  plan is prepared in its energy ground state
and interacts with a laser beam of a Laguerre-Gaussian form. The
expectation value of the angular momentum in the long time limit
\cite{Loud} shows two distinct resonances at $\omega_a, \omega_b.$
In an appropriate laser trapping field the speed of the atom can
be slowed to the extent that the kinetic energy term in
(\ref{Eq:CS-H1}) may be removed \cite{SST}. As the kinetic energy
diminished, only one resonance remains at $\omega^{\ast}=K/G,$ and
the dominate value of the corresponding lowest angular momentum is
$\hbar/2.$ Furthermore, as the magnetic field eliminated, the
parameter $g$ approaches zero, the resonance occurs at $\tilde
\omega=\tilde K/\tilde G\sim \hbar/(\mu\theta),$ and the dominate
value of the corresponding lowest angular momentum shifts to
$\hbar/4.$
Since a Laguerre-Gaussian beam carries orbital angular momentum
along its direction of propagation \cite{ABSW}, an atom moving in
such a beam is subject to a radiation-induced torque, which is
proportional to the eigenvalue of mode's orbital angular momentum
\cite{BPA}. This suggests that a Laguerre-Gaussian beam supplies a
suitable probe for the above angular momentum resonances.

Of course, any attempt to detect effects of spatial
noncommutativity, if any, is a challenging enterprise. In the
above case the dominate value of the lowest angular momentum
$\hbar/4$ is independent of the parameter $\theta,$ but the
frequency $\tilde \omega$ is $\theta$ dependent. There are
different bounds on the parameter $\theta$ set by experiments. The
space-space noncommutative theory from string theory violates
Lorentz symmetry and therefore strong bounds can be placed on the
parameter $\theta$, the existing experiments \cite{CHKLO} give
$\theta/(\hbar c)^2\le (10 \;TeV)^{-2}$. Comparing with the above
estimation, other bounds on $\theta$ exist: measurements of the
Lamb shift \cite{CST} give a weaker bound;
clock-comparison experiments \cite{MPR} claim a stronger
bound.
The magnitude of $\theta$ is surely extremely small; the frequency
$\tilde \omega$ is surely extremely large.
\footnote{\;The dominate value of the frequency $\tilde \omega$ is
$\tilde \omega = \tilde K/\tilde G\approx \hbar/(2\mu\theta)$. If
we take $\mu c^2=2 \,GeV$ and $\theta/(\hbar c)^2\le (10^4
\;GeV)^{-2}$ we obtain $\tilde \omega \ge 10^{32} Hz$.}

\vspace{0.4cm}

This work has been partly supported by the National Natural
Science Foundation of China under the grant number 10074014 and by
the Shanghai Education Development Foundation.

\end{document}